\documentclass{aastex62}

\usepackage{lineno}

\newcommand\aastex{AAS\TeX}

\received{2018 September 28}
\revised{2018 October 17}
\accepted{2018 October 17}

\submitjournal{ApJ}

\shorttitle{\aastex \ Type II burst band-splitting}
\shortauthors{Chrysaphi et al.}

\graphicspath{{figures/}}

\begin{document}

\title{CME-DRIVEN SHOCK AND TYPE II SOLAR RADIO BURST BAND-SPLITTING}
\correspondingauthor{Nicolina Chrysaphi}
\email{n.chrysaphi.1@research.gla.ac.uk}

\author[0000-0002-4389-5540]{Nicolina Chrysaphi}
\affiliation{School of Physics \& Astronomy, University of Glasgow, Glasgow, G12 8QQ, UK}

\author[0000-0002-8078-0902]{Eduard P. Kontar}
\affiliation{School of Physics \& Astronomy, University of Glasgow, Glasgow, G12 8QQ, UK}
\collaboration{}

\author[0000-0002-2219-100X]{Gordon D. Holman}
\affiliation{NASA Goddard Space Flight Center, Code 671, Greenbelt, MD 20771, USA}
\collaboration{}

\author[0000-0003-4867-7558]{Manuela Temmer}
\affiliation{Institute of Physics, University of Graz, 8010 Graz, Austria}
\collaboration{}

\begin{abstract}
Coronal Mass Ejections (CMEs) are believed to be effective in producing shocks in the solar corona and the interplanetary space.  One of the important signatures of shocks and shock acceleration are Type II solar radio bursts that drift with the shock speed and produce bands of fundamental and higher harmonic plasma radio emission.  An intriguing aspect of Type II radio bursts is the occasional split of a harmonic band into thinner lanes, known as band-splitting.  Here, we report a detailed imaging and spectroscopic observation of a CME-driven shock producing band-splitting in a Type II burst. Using the Low Frequency Array (LOFAR), we examine the spatial and temporal relation of the Type II burst to the associated CME event, use source imaging to calculate the apparent coronal density, and demonstrate how source imaging can be used to estimate projection effects.  We consider two widely accepted band-splitting models that make opposing predictions regarding the locations of the true emission sources with respect to the shock front.  Our observations suggest that the locations of the upper and lower sub-band sources are spatially separated by $\sim 0.2 \pm 0.05 \, \mathrm{R_\sun}$.  However, we quantitatively show, for the first time, that such separation is consistent with radio-wave scattering of plasma radio emission from a single region, implying that the split-band Type II sources could originate from nearly co-spatial locations.  Considering the effects of scattering, the observations provide supporting evidence for the model that interprets the band-splitting as emission originating in the upstream and downstream regions of the shock front, two virtually co-spatial areas.
\end{abstract}


\section{Introduction} \label{section_intro}
Coronal Mass Ejections (CMEs) produce a variety of radio signatures associated with non-thermal electrons. Type II solar radio bursts are the bright radio emissions often associated with CMEs and characterized by a slow frequency drift rate ($\lesssim -1 \, \mathrm{MHz \, s^{-1}}$) across the dynamic spectrum \citep{1985srph.book.....M}. It is believed that they are excited by shock waves and are thus closely linked to solar eruptive events \citep{1964SvA.....7..639P, 1999GeoRL..26.2331D,2000ApJ...528L..49M, 2006SSRv..123..341P, 2010ApJ...712.1410T, 2011SoPh..273..433G, 2011SoPh..273..143V, 2017SoPh..292..161K, 2018ApJ...863L..39G, 2018A&A...615A..89Z}.  Type II radio emission likely originates in enhanced density regions of the corona \citep{2003ApJ...590..533R} - possibly in coronal streamers - and sometimes demonstrates two characteristic bands with a frequency ratio of approximately 1:2.  These bands are associated with plasma emission near the local plasma frequency (fundamental) and its second harmonic \citep{1950AuSRA...3..387W,1985srph.book.....M}.  Occasionally, each of these bands will further split into two (or more) thinner lanes approximately parallel to each other \citep{1985srph.book.....M}.  This phenomenon is known as ``band-splitting" and although multiple theories proposed over the past decades have attempted to explain the underlying process, the mechanism generating this emission is still debated.  An interesting characteristic of split-band Type II bursts is the observed frequency split $\Delta f/f$ between sub-bands which tends to remain nearly constant over the course of a single event, as well as from one event to another \citep{2001A&A...377..321V, 2015ApJ...812...52D}.  This frequency split $\Delta f/f$ is found to most often range between 0.1 and 0.5 (see e.g. Figure 5 of \citet{2001A&A...377..321V}), and despite the various interpretations addressing this band-splitting range, it is still unclear what determines these values.

Of the plethora of band-splitting models proposed in the literature, we consider the two most widely accepted interpretations, both of which explain the observed emission in terms of geometrical effects but make opposing predictions regarding the spatial origins of the true emission sources.

\citet{1974IAUS...57..389S, 1975ApL....16R..23S} proposed a model that attributes the splitting of the bands as the emission from the upstream (ahead) and downstream (behind) regions of the shock front.  By adopting this interpretation, the Rankine-Hugoniot jump conditions \citep{2014masu.book.....P} can be invoked through the relation of the frequency ratio of the sub-bands to the density jump across the shock front \citep{1995A&A...295..775M}.  Once the shock speed is calculated, the jump conditions enable the estimation of the Alfv\'en Mach number, the Alfv\'en speed, and consequently the local coronal magnetic field strength \citep{2002A&A...396..673V}. In this scenario, the radio sources near the Sun should be virtually co-spatial as the shock's thickness is negligible.

\citet{1983ApJ...267..837H} proposed a model explaining the band-splitting as radiation produced by electrons reflected upstream (ahead) of the shock front at regions where the curvature of the shock is quasi-perpendicular to the coronal magnetic field.  Unlike the \citet{1974IAUS...57..389S, 1975ApL....16R..23S} model, the emission producing each sub-band originates from different parts of the shock front, meaning that the radio sources are expected to be spatially separated.

Intriguingly, spatially resolved observations conducted with the \textit{Nan\c{c}ay Radioheliograph} \citep[NRH;][]{1997LNP...483..192K} for a limited number of frequencies demonstrated spatial separation between the sources of split-band Type II bursts \citep{2012A&A...547A...6Z,2015AdSpR..56.2811Z}.  Only sparsely separated frequencies ($20-50$ MHz separation) could be observed simultaneously with NRH, making the interpretation more difficult given that the width of a Type II band is around 30 MHz for emissions near 150 MHz.  Therefore, simultaneous observations at multiple frequencies within the split band are needed to test the band-splitting models.

In this paper, we present a Type II radio burst observed by the \textit{LOw-Frequency ARray} \citep[LOFAR;][]{2013A&A...556A...2V} which provides us with unprecedented imaging capabilities and allows for a more detailed analysis of radio emissions.  LOFAR consists of two main types of antennas: the Low-Band Antenna (LBA) composed of dipoles, and the High-Band Antenna (HBA) composed of tiles, and collectively it covers a frequency range of $10-240$ MHz.  Individual antennas can be grouped together to form beams that can later be computationally phased.  A coherent summation of the beams will result in what is referred to as a ``tied array beam" \citep{2013A&A...556A...2V}.  This observing mode is the ideal mode for observations of solar radio emissions as it allows for high temporal, spectral, and spatial resolution, all necessary to capture the rapidly changing, small-scale variations of radio sources that can often expand over areas comparable to the Sun itself as they move away from the limb \citep{2017A&A...606A.141R}.

LOFAR's unprecedented computational capability enables it to image the emission source at every $\sim 10$~ms (temporal cadence) and $\sim 12$~kHz (spectral sampling) near 30 MHz.  We are therefore able to show, for the first time, emission source images of both structures of a split-band Type II burst at exactly the same moment in time.  This allows for a comparison of the spatial relations of the upper and lower sub-band sources without any time delay ambiguities in the observations.  LOFAR's imaging capability represents a significant improvement over previous instruments capable of source imaging, like the NRH which could image at up to 10 non-consecutive frequencies between 150 and 450 MHz, thus preventing the simultaneous observation of both the upper and lower sub-band sources \citep{2012A&A...547A...6Z}.

The paper is structured as follows: section \ref{section:observations} describes how the observations of the CME, the activities on the solar surface, and the radio emission were conducted.  In section \ref{section:CME_analysis} we obtain estimations on the CME's expansion and speed, and examine the temporal and spatial relation of the CME and the Type II emission.  We then use LOFAR imaging to calculate an average spatial separation of $ \sim 0.2 \pm 0.05\, \mathrm{R_\sun}$ between the higher and lower frequency components of the split-band Type II.  In section \ref{section:obs_vs_model_R} we assume a coronal density model to compare model-predicted source locations at different frequencies to the observed locations at equivalent frequencies.  We find that the density model needs to be multiplied by a factor of 4.5 to match the heights of the observed source locations.  In section \ref{section:out-of-plane_R} we present a way to estimate the out-of-plane heliocentric distances of the sources, and in section \ref{section:scattering} we quantitatively estimate the effect of scattering on split-band Type II sources.  We show that the spatial separation is consistent with radio-wave scattering effects, meaning that the true sources of the two sub-bands can be virtually co-spatial.  We also find that the radial shift of the source location caused by scattering will make the density appear 4.3 times larger than its true value.  A discussion and summary of our results is presented in section \ref{section:discussion}.

\section{Overview of the observations} \label{section:observations}
We study the radio signatures associated with a CME event on 2015 June 25 near 10:57 UT.  Coronagraphic white-light images of the CME were captured by the \textit{Solar and Heliospheric Observatory} \citep[SOHO;][]{1995SSRv...72...81D} Large Angle Spectroscopic Coronagraph \citep[LASCO;][]{1995SoPh..162..357B}.  The eruption could be observed and measured within the field-of-view of LASCO's C2 coronagraph which covers a distance range of $1.5-6 \, \mathrm{R_\sun}$, but can only image beyond $2.2 \, \mathrm{R_\sun}$.  Beyond that range, the CME dissolves into the background which is dominated by an earlier CME event.  Observations from the \textit{Solar Dynamics Observatory} \citep[SDO;][]{2012SoPh..275....3P} Atmospheric Imaging Assembly \citep[AIA;][]{2012SoPh..275...17L} are also used to identify the origin of the eruption with respect to activities on the solar surface.

In close spatial and temporal proximity to the CME's C2 appearance, a solar radio burst identified as a Type II burst was recorded at $\sim$ 10:46 UT by LOFAR.  As Type II bursts are thought to be related to the shock waves caused by solar eruptive events, the relation of the burst and the associated CME is probed.

The 2015 June 25 Type II observation was conducted using LOFAR's outer LBA core stations and utilising the Coherent Stokes beam-formed mode.  This imaging scheme produced a tied-array beam of 171 individual beams covering a hexagonal area extending up to $ \sim 2 \, \mathrm{R_\sun}$ from the solar center.  The solar observation was calibrated using Tau A, as described in \citet{2017NatCo...8.1515K} and \citet{2018SoPh..293..115S}.

The Type II burst was observed by LOFAR for 4128 consecutive frequencies across a bandwidth of 50.4 MHz between 30 and 80 MHz, with a spectral and temporal resolution of approximately 12.2 kHz and 0.01 seconds, respectively.  A brighter than the background feature that slowly drifts to lower frequencies as time increases (by $\sim -0.1 \, \mathrm{MHz \, s^{-1}}$) is observed - a characteristic behavior of Type II bursts.  Splitting of this feature in two thinner lanes is also observed between 10:46:00 and 10:48:00 UT defining the band-splitting region recorded.  Type III radio bursts identified by their high frequency drift rates and short lifetimes were also observed during the Type II observation, signaling the presence of open magnetic field lines.

\section{Results} \label{section:results}
\subsection{The CME--Type II relation} \label{section:CME_analysis}
\begin{figure}[ht!]
    \centering
    \includegraphics[width=0.85\textwidth, keepaspectratio=true]{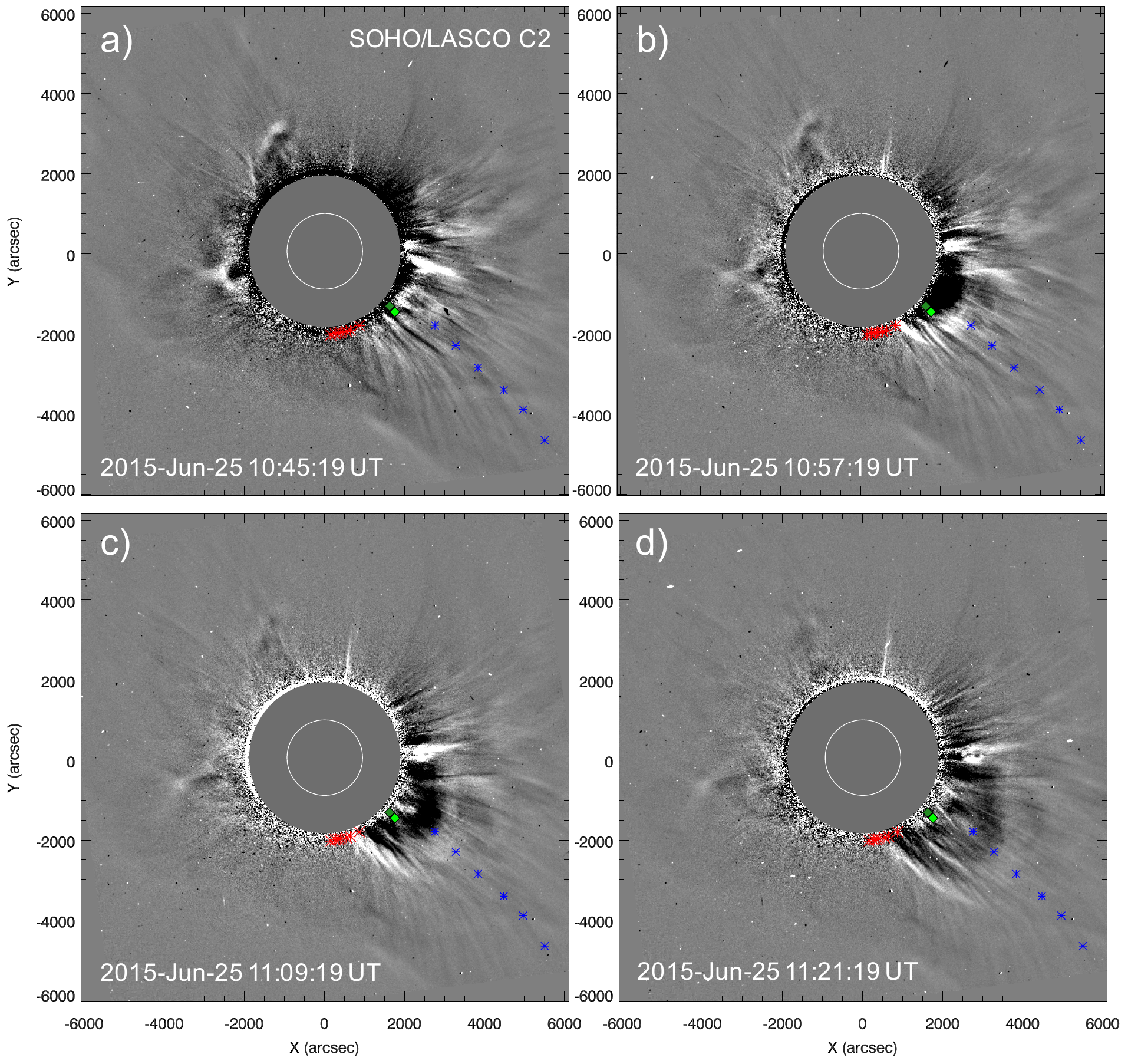}
    \caption{Panels (a)--(d) illustrate the way in which consecutive coronagraph images captured by LASCO/C2 ($\sim$ 12 minutes apart) were used to track different features of the CME.  Blue stars indicate the tracking of the CME's front and red stars indicate the CME's lateral expansion.  The dark and light green diamonds represent the apparent location of the Type II upper and lower sub-bands, respectively, at 10:46:29 UT.
    }
    \label{fig:cme_tracking_annot}
\end{figure}

A CME event was observed by the LASCO C2 coronagraph emerging from the south-west part of the solar limb.  The eruption first appears in the C2 field-of-view  at $\sim$ 10:57 UT and gradually fades into the highly disturbed background dominated by the residual structures of an earlier CME event that was observed by C2 at $\sim$ 8:36 UT.  Near the time of the CME appearance in C2, radio emissions were recorded by LOFAR between $30-80$ MHz.  Signatures of a Type II radio burst were identified from $\sim$ 10:46 UT.  Given that the exciters of Type II bursts are known to be shock waves often associated with CMEs, we study the spatial and temporal characteristics of the CME with relation to the Type II burst.  We use consecutive running difference images to track the expansion of the CME's front, as well as the CME's lateral expansion, and estimate the propagation speed in both directions.  The CME features used to obtain these estimates, along with the apparent location of the split-band Type II burst at 10:46:29 UT, are indicated in Figure \ref{fig:cme_tracking_annot}.

\begin{figure*}[ht!]
    \centering
    \includegraphics[width=1\textwidth, keepaspectratio=true]{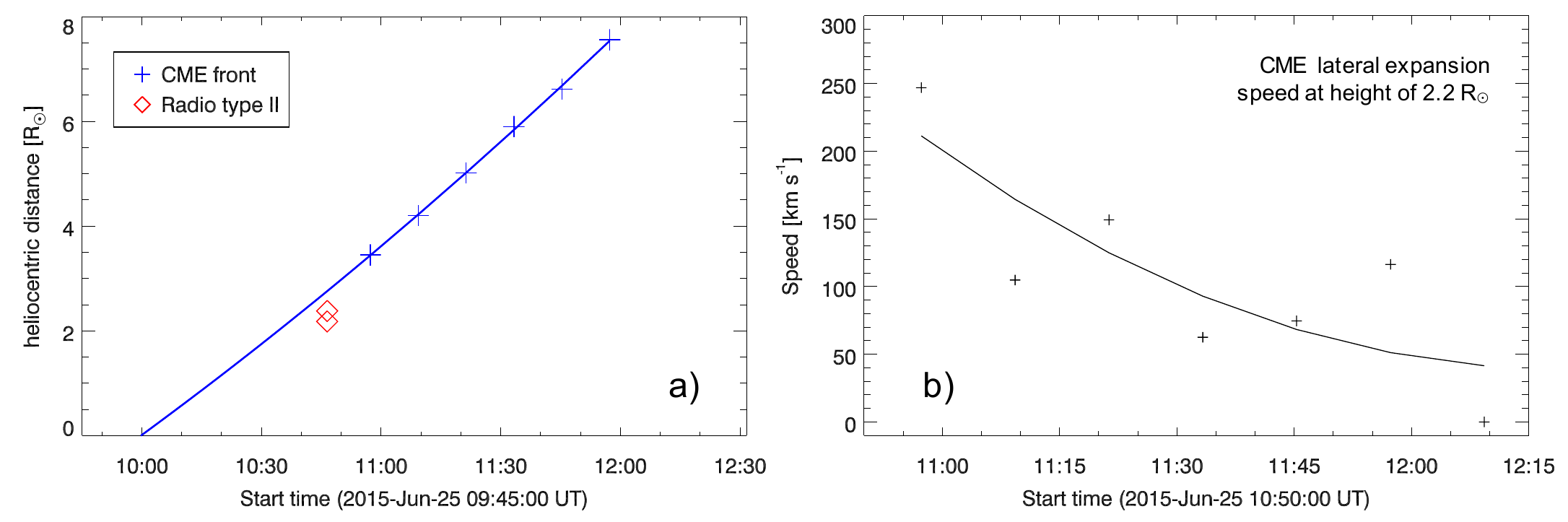}
    \caption{(a) CME height as a function of time obtained by tracking the CME's front (blue crosses) using LASCO/C2 images as illustrated in Figure \ref{fig:cme_tracking_annot}.  The red diamonds represent the apparent location of the Type II sub-bands at 10:46:29 UT.  The blue line represents a fit through the tracked trajectory of the CME's front which was extrapolated in order to approximate the height of the CME at the time of the radio emissions, as well as the CME's eruption time.  The mean CME speed was derived to be approximately $740 \, \mathrm{km \, s^{-1}}$.
    (b) An estimation of the speed with which the CME expanded in the lateral direction at a height of $2.2 \, \mathrm{R_\sun}$, as obtained from the analysis of LASCO/C2 running difference images illustrated in Figure \ref{fig:cme_tracking_annot}.
    }
    \label{fig:cme_fits}
\end{figure*}

The left panel of Figure \ref{fig:cme_fits} illustrates the CME's propagation in the radial direction and gives the heliocentric distance of the CME front as a function of time.  A non-linear fit through the frontal CME features (blue crosses) is used to derive a mean CME speed over the LASCO C2 field-of-view of approximately $740 \, \mathrm{km \, s^{-1}}$.  Given that SDO observations do not provide a clear signature of the time the CME launches, and no impulsive phase can be distinguished in X-ray measurements since a flare at $\sim$ 9:00 UT masks the following eruptions, a back-extrapolation is needed to estimate the CME's eruption time.  The back-extrapolation of the fit (Figure \ref{fig:cme_fits}) places an estimate on the CME eruption at $\sim$ 10:15 UT and indicates that the CME was at a height of $\gtrsim 2.5 \, \mathrm{R_\sun}$ above the solar center when the Type II burst was first detected.  The lateral expansion of the CME flank is measured at a constant height of $2.2 \, \mathrm{ R_\sun}$, and from the distance-time measurements we derive the expansion speed over time, as shown in the right panel of Figure \ref{fig:cme_fits}.  The overall trend of the lateral expansion shows a deceleration with progressing time.

\begin{figure*}[ht!]
    \centering
    \gridline{\hspace{-1.3cm}
              \leftfig{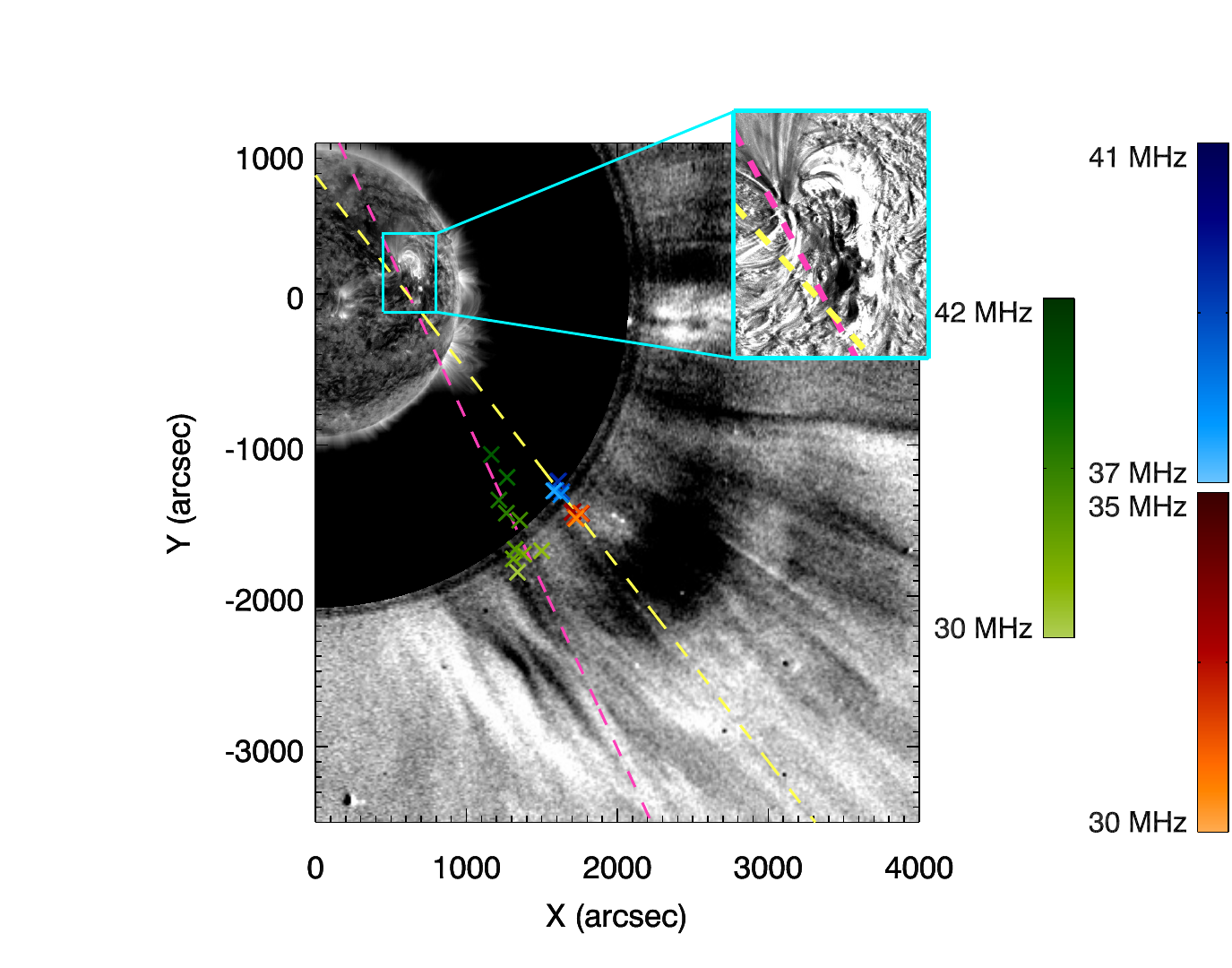}{0.55\textwidth}{(a)}
              \hspace{-1.2cm}
              \rightfig{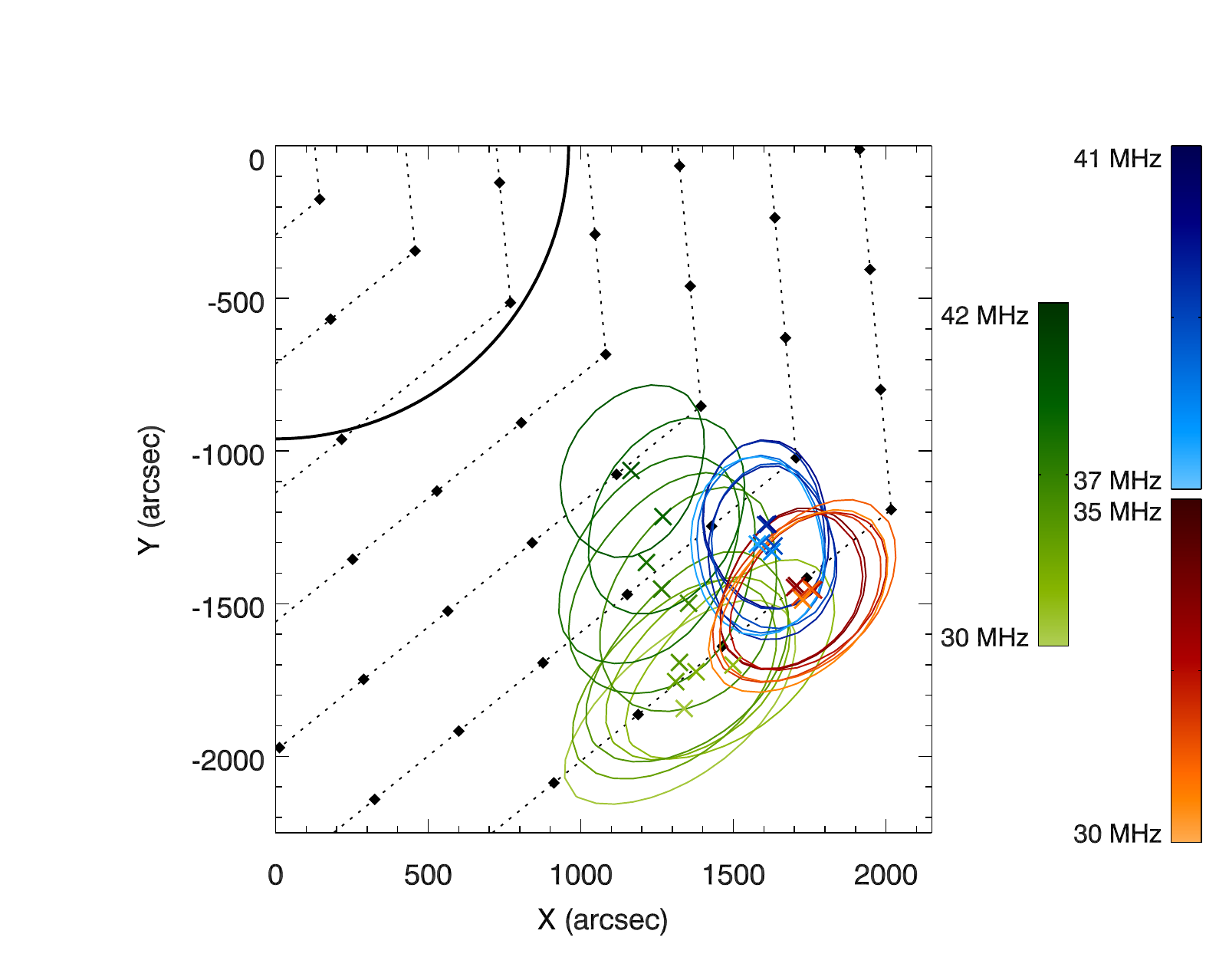}{0.55\textwidth}{(b)}
              }
    \caption{(a) A combination of data from SDO/AIA at 171\AA, SOHO/LASCO/C2, and LOFAR.  The centroids in blue color scheme represent the upper sub-band sources, and the ones in red represent the lower sub-band sources.  The green centroids represent the Type III burst occurring at 10:47:43 UT (see Figure \ref{fig:dyn_spec}), whereas the color gradients represent movement in frequency with darker colors indicating higher frequencies.  Linear fits (yellow and magenta dashed lines) were applied through the Type II and Type III centroids (respectively), with both fits appearing to point back towards the active region.  Running difference images from LASCO/C2 are used to highlight the structures of the CME and the coronal streamer which seem to spatially relate to the Type II and Type III burst, respectively.  The inset is a running ratio image of SDO/AIA observations at 193\AA.  It indicates the area below the active region where the dimming is observed, and from which the CME is though to have originated, while also highlighting that the fits through the radio sources (yellow and magenta lines) point back towards this region.
    (b) The centroid locations and respective 90\% maximum intensity contours for the Type III (green) and the upper (blue) and lower (red) Type II sub-bands with respect to the solar limb (solid black curve).  The color scheme represents movement in frequency, with darker colors corresponding to higher frequencies.  The black diamonds show the central locations of the LOFAR beams, and in collective the Field-of-View of the tied-array beam.
    }
    \label{fig:centroid_fit_and_contours}
\end{figure*}

To locate the origin of the CME seen at $\sim$ 10:57 UT, SDO/AIA images of the solar surface close to the time of the CME and Type II burst were studied.  AIA imaging shows that at $\sim$ 9:50 UT there is considerable dimming of the Southern part of the active region from which the CME observed at $\sim$ 8:36 UT originated, signaling density depletion and mass loss related to the ejection of CMEs.  The time of the dimming agrees with our estimate of the CME eruption at $\sim$ 10:15 UT.  This leads us to believe that the CME observed at $\sim$ 10:57 UT is most likely to have originated from the region of the dimming \citep{1996ApJ...470..629H, 2018ApJ...863..169D}.  This is also the region towards which a fit through all Type II and Type III centroids point (cf. Figure \ref{fig:centroid_fit_and_contours}).

\begin{figure}[ht!]
    \centering
    \includegraphics[width=0.95\textwidth, keepaspectratio=true]{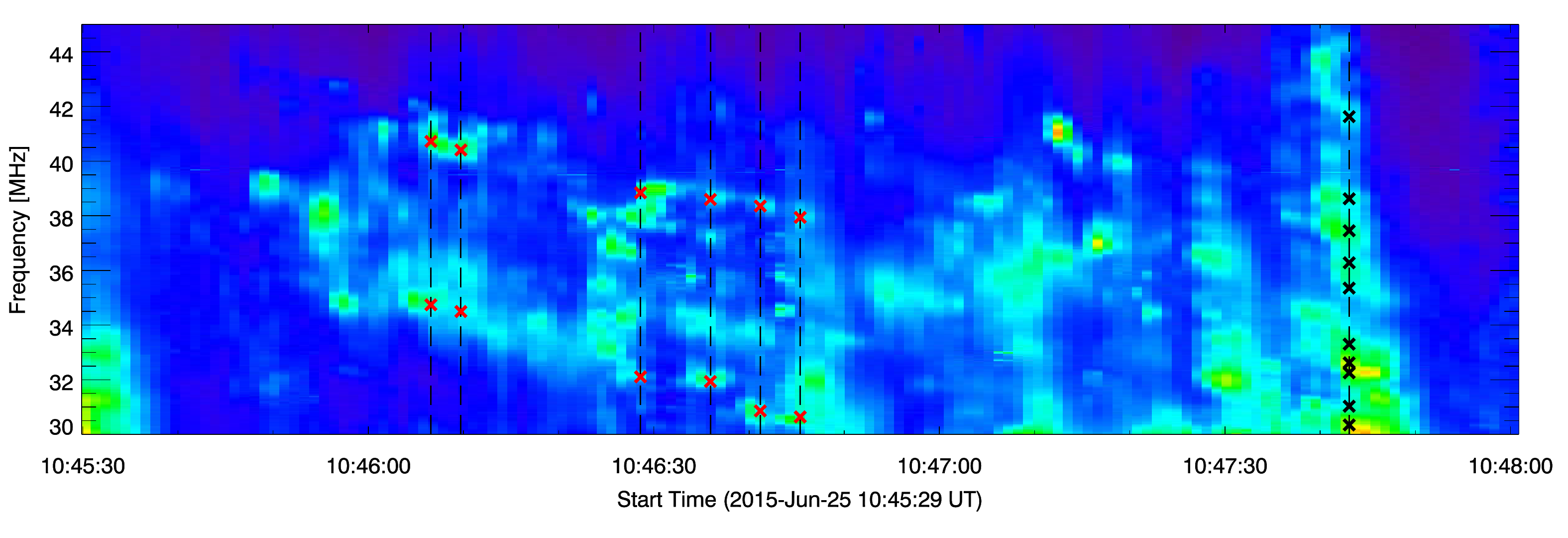}
    \caption{Dynamic spectrum depicting part of the radio emissions observed on 2015 June 25 by LOFAR's LBA between 10:45:30 and 10:48:00 UT.  The temporal and spectral resolutions were re-binned prior to plotting decreasing them to 24.4 kHz and 1 second, respectively.  Black dashed lines indicate moments of a single time, whereas crosses represent the moments in time and frequency for which the emission images were used to estimate the centroid locations of the sources.  A point from the higher frequency and a point from the lower frequency Type II sub-band (red crosses) was selected for each of the six moments in time.  Similarly, points along the Type III burst (black crosses) were selected to represent the Type III source motion in frequency.
	}
    \label{fig:dyn_spec}
\end{figure}

Of particular interest are the origins of the Type II sub-band sources with respect to the CME, so to tackle this question we temporally and spatially relate the locations of the Type II sources to the CME structures.  Observations of different wavelengths and fields of view from a combination of spacecraft were used in order to image the variety of events, as illustrated by Figure \ref{fig:centroid_fit_and_contours} where the color schemes represent motion in frequency of the radio sources.  To represent the split-band Type II sources we select points from each of the high and low frequency sub-bands at six moments in time (c.f. red crosses on Figure \ref{fig:dyn_spec}), and estimate the centroid location of the radio emission sources.  The estimation of the source centroids is accomplished by fitting a 2-dimensional (2D) elliptical Gaussian to the LOFAR images.  These centroids are indicators of the position of a Type II source for a given time and frequency, and allow us to compare the radio sources against the CME and background coronal and solar structures (left panel of Figure \ref{fig:centroid_fit_and_contours}).  It can be seen that the Type II sources appear to originate at the Southern flank of the CME where potential compression between the CME and the streamer could have taken place.  The streamer arose during the interactions of the CME event at $\sim$ 8:36 UT and is traced by Type III sources which are indicators of the existence of open magnetic field lines.  The heights at which the Type II appears are not covered by the C2 coronagraph, and other coronagraphic observations from instruments imaging smaller heliocentric distances are unavailable for the studied event.  Due to these observational limitations, we are unable to determine the origins of the Type II with respect to the CME structures with complete certainty.  However, the similarity in the morphology and evolution of the two sub-bands (see Figure \ref{fig:dyn_spec}) suggests that the Langmuir waves producing the split-band radio emission are excited by the same coronal structures.

Using the obtained centroid positions, the spatial separation between each higher and lower frequency sub-band source at each of the six moments in time (cf. Figure \ref{fig:dyn_spec}) was calculated.  The average emission source separation of the Type II upper and lower sub-bands, as seen in Figure \ref{fig:centroid_fit_and_contours}, is approximately $0.2 \pm 0.05 \, \mathrm{R_\sun}$.

\subsection{Model-predicted Source Locations} \label{section:obs_vs_model_R}
As Type II radio bursts are the result of plasma emission, the observed radiation roughly corresponds to the local fundamental plasma frequency or a higher harmonic of it.  Assuming fundamental emission ($f_{observed}=f_{pe}$), the electron plasma frequency $f_{pe}$ is related to the density by
$ f_{pe} \, \mathrm{[Hz]} = \kappa \sqrt{n_{e}} , $
where
$ \kappa = \sqrt{e^{2}/\pi m_{e}} $,
$n_e$ is the electron plasma density in $\mathrm{cm^{-3}}$, $e$ is the electron charge, and $m_e$ is the electron mass.

Consequently, the plasma density can be estimated for any frequency, and by taking a coronal density model the heliocentric distance for each frequency can be obtained.  Using, in this case, the standard \citet{1961ApJ...133..983N} coronal density model,
\begin{equation} \label{eqn:ne}
n_{e} = N \cdot n_0 \cdot 10^{4.32R_{\sun}/R}	\quad \mathrm{[cm^{-3}]} \,,
\end{equation}
where $N$ is the constant denoting the multiplicative factor of the density model, and $ n_0=4.2 \times 10^{4} \, \mathrm{cm^{-3}} $, the distance from the solar center $R$ can be written as:
\begin{equation} \label{eqn:R_model}
\frac{R}{R_\sun} = \frac{2.16}{\log_{10}(f_{pe})-\log_{10}(\kappa \cdot \sqrt{n_{0}N})} \,.
\end{equation}
One can then estimate the radial speed of the shock wave as a function of distance,
\begin{equation} \label{eqn:Vshock}
V_{shock} = \frac{2}{f_{pe}} \frac{df_{pe}}{dt} \left(\frac{d}{dR} \ln(n_{e})\right)^{-1} \,,
\end{equation}
where $df_{pe}/dt$ is the frequency drift rate in $\mathrm{Hz \, s^{-1}}$ obtained from the dynamic spectrum, and $(d \ln (n_{e})/dR)^{-1} = -R^2 / (4.32 R_\sun \ln(10))$ for the Newkirk density model.

\begin{figure}[ht!]
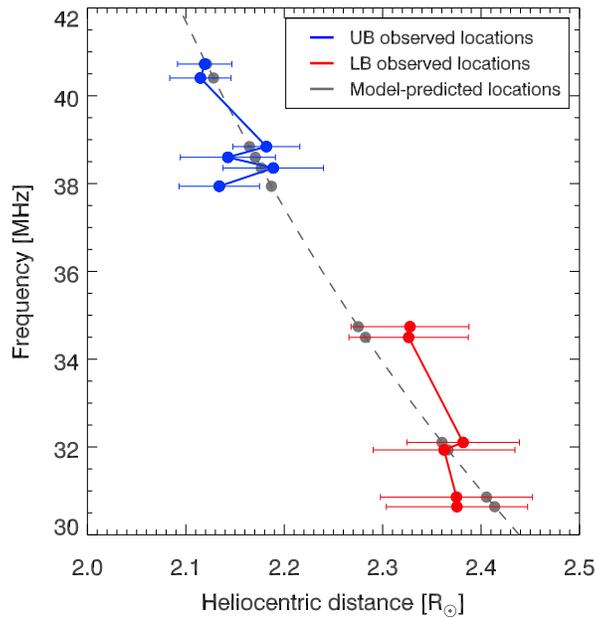

    \centering
    \includegraphics[width=0.45\textwidth, keepaspectratio=true]{{{4.5xN_vs_obs}}}    
    \caption{A comparison of the observed heliocentric centroid locations (as estimated by the 2D elliptical Gaussian fit) and the source locations predicted by the 4.5$\times$Newkirk density model (gray dashed line).  Apparent upper (UB) and lower (LB) sub-band centroids are indicated in blue and red colored circles, respectively, whereas model-predicted locations for the equivalent frequencies are indicated in gray circles.  The errors on the source positions were estimated using the equations presented in \citet{2017NatCo...8.1515K}.
    }
    \label{fig:4.5N_vs_obs}
\end{figure}

The model-predicted heliocentric distance $R$ of a source at a specific frequency (as obtained from Equation (\ref{eqn:R_model})) can be compared to the observed heliocentric distance of a source (as obtained by the centroid estimations directly from images) at an equivalent frequency.  This is shown in Figure \ref{fig:4.5N_vs_obs} where a multiplicative factor N (see Equation (\ref{eqn:ne})) of 4.5 was found to produce model-predicted locations that best match the observed values (see \citet{2010ApJ...718..266M} for a similar procedure using NRH data).

It should be noted that, although comparing the model-predicted distances to the observed source distances is a better approach than randomly selecting a density model multiplicative factor $N$ to describe the corona's profile, it is nevertheless non-ideal.  The observed emission sources not only suffer from projection effects but also from radio-wave propagation effects, the dominant of which is scattering as a recent investigation by \citet{2017NatCo...8.1515K} has shown for the LOFAR frequency range.

\subsection{Estimation of Projection Effects} \label{section:out-of-plane_R}
Images of solar eruptions are affected by line-of-sight (LoS) effects which distort the true 3-dimensional location of sources during the translation to the 2-dimensional plane-of-sky depiction.  Similarly, the apparent locations of the Type II sources as viewed in the plane of the sky are likely to result in underestimated heliocentric distances.
An estimation of the out-of-plane heliocentric distance of the sources can be obtained due to LOFAR's ability to observe the sources of both sub-bands at any time and any frequency.  This however is only possible given the assumption that both sources are within the same atmosphere, i.e. the same density model determines the location of both sources.  Such assumption is valid for an interpretation like the one proposed by \citet{1983ApJ...267..837H} where both sources are produced upstream of the shock, 
but not for the model proposed by \citet{1974IAUS...57..389S, 1975ApL....16R..23S}.

Given a density model - in this case the \citet{1961ApJ...133..983N} model - an expression for the ratio of the heliocentric distance (see Equation (\ref{eqn:R_model})) of two sources of given frequencies 
can be obtained:
\begin{equation} \label{eqn:R_ratio}
\frac{R_L}{R_U} = 1 - \frac{2}{4.32} \cdot R_L \cdot \log_{10}{\left(\frac{f_L}{f_U}\right)} \,.
\end{equation}
In this equation, $R_L$ and $R_U$ represent the out-of-plane heliocentric distances ($R_{H_{out}}$, c.f. Equation (\ref{eqn:R_Hout})) of the lower and upper sub-band sources, respectively, and $f_L$ and $f_U$ are the frequencies at which each source was observed, with $f_U$ being the higher frequency component.

Expressions for the out-of-plane heliocentric distances can be obtained through geometric relations, with known parameters directly acquired from the images.  The out-of-plane heliocentric distance $R_{H_{out}}$ of a source is a function of the in-plane heliocentric distance $R_{H_{in}}$ of that source, and a certain height $z$ out of the plane:
\begin{equation} \label{eqn:R_Hout}
R_{H_{out}} = \sqrt{z^2 + R_{H_{in}}^2} \,.
\end{equation}
The out-of-plane height $z$ of the source however, is an unknown parameter.  An assumption that both sources follow the same trajectory away from the solar surface, and in this case the active region AR (see Figure \ref{fig:centroid_fit_and_contours}), can be made in order to obtain an expression for the out-of-plane height $z$:
\begin{equation} \label{eqn:z}
z = R_{AR_{in}} \cdot \tan{\theta_{AR}} \,.
\end{equation}
The quantity of $R_{H_{in}}$, and hence $R_{AR_{in}}$ (the in-plane distance of the source from the AR), can be estimated using the observation images.  To select a starting point from which to estimate the distance of the sources from the AR, we consider a point that is along the linear fit through the Type II centroids representing their path (see Figure \ref{fig:centroid_fit_and_contours}), and within the area of dimming believed to be the origin of the CME (see section \ref{section:CME_analysis}).  The angle $\theta_{AR}$ is the angle between the active region and the out-of-plane height $z$, and it is an unknown parameter.  It can, however, be computed using the relation given by Equation (\ref{eqn:R_ratio}) as the solution will be the value of angle $\theta_{AR}$ to satisfy the equality.

One can then compare the model-predicted heliocentric distances (see Equation (\ref{eqn:R_model})) to the estimated out-of-plane heliocentric distances and obtain a new value for the density model multiplicative factor $N$ that matches the out-of-plane estimations (see section \ref{section:obs_vs_model_R} and Figure \ref{fig:4.5N_vs_obs}).  Once the value of $N$ is acquired, the local coronal density can be estimated (see Equation (\ref{eqn:ne})) using the obtained out-of-plane heliocentric locations of the sources.

The observational limitations of the Type II burst examined here do not allow us to compute the out-of-plane distances with confidence.  Future split-band Type II observations conducted with instruments of imaging capabilities similar to LOFAR can, however, be used to test the mathematical model presented.

\subsection{Estimation of Scattering Effects} \label{section:scattering}
A recent investigation into observations of solar radio emissions by \citet{2017NatCo...8.1515K} showed that sources with small intrinsic sizes of $\sim 0.1$ arcmins are observed as large as $\sim 20$ arcmins.  The observations suggest that the source sizes and positions at fundamental emission are determined by radio-wave propagation effects, the dominant of which is scattering of the radio waves.
In this section we present, for the first time, a quantitative estimation of the effect that scattering can have on the apparent source locations and the relative separation of the higher and lower frequency components of a split-band Type II burst.

Following \citet{1952MNRAS.112..475C}, \citet{1968AJ.....73..972H}, \citet{1971A&A....10..362S}, \citet{1999A&A...351.1165A}, and \citet{2007ApJ...671..894T},
we consider homogeneous and stationary density fluctuations with a spatial autocorrelation function
\begin{equation}\label{eqn:dn_dn}
  \langle\delta n(\vec{r}_1)\delta n(\vec{r}_2)\rangle=
  \langle\delta n^2\rangle\exp\left(-\frac{(\vec{r}_1-\vec{r}_2)^2}{h^2}\right),
\end{equation}
where $h$ is the characteristic density inhomogeneity correlation scale,
and $\langle ... \rangle$ denotes an ensemble average.

Isotropic density inhomogeneities (Equation (\ref{eqn:dn_dn})) cause radio waves of frequency $f$ to experience an angular scattering.  The expression for angular scattering (which can be derived using e.g. Equations (3), (6), and (7) from \citet{1971A&A....10..362S,2018ApJ...857...82K}) is given as:
\begin{equation}\label{eqn:dtheta}
  \frac{d\langle\Delta \theta^2\rangle}{dr}
  =\frac{\sqrt{\pi}}{2h}\frac{f_{pe}^4(r)}{(f^2-f^2_{pe}(r))^2}\frac{\langle\delta n^2\rangle}{n^2} \,,
\end{equation}
where $f^2_{pe}=\omega_{pe}^2/(2\pi)^2= e^2 n / \pi m$ is the electron plasma frequency and $n$ is the electron plasma density. Equation (\ref{eqn:dtheta}) indicates that the scattering rate depends on ${\langle\delta n^2\rangle}/({n^2}h)$.  It is also a decreasing function of radial distance, meaning that when $f$ is close to $f_{pe}$ the scattering is frequent, whereas at large distances from the Sun, where $f \gg f_{pe}$, the scattering becomes negligible.  To quantitatively characterize the effect of wave scattering, the optical depth with respect to the scattering is given by:
\begin{equation}\label{eqn:tau}
   \tau(r)=\int_{r}^{1AU} \frac{d\langle\Delta \theta^2\rangle}{dr}dr=
   \int_{r}^{1AU} \frac{\sqrt{\pi}}{2}\frac{f_{pe}^4(r)}{(f^2-f^2_{pe}(r))^2} \frac{\epsilon^2}{h} dr \,.
\end{equation}
One can assume that the distance at which the radio-wave optical depth (with respect to scattering), $\tau = 1$ is defined as the distance at which the transition between a region of strong scattering and a region of weak/no scattering occurs.

\begin{figure}[ht!]
    \centering
    \includegraphics[width=0.55\textwidth, keepaspectratio=true]{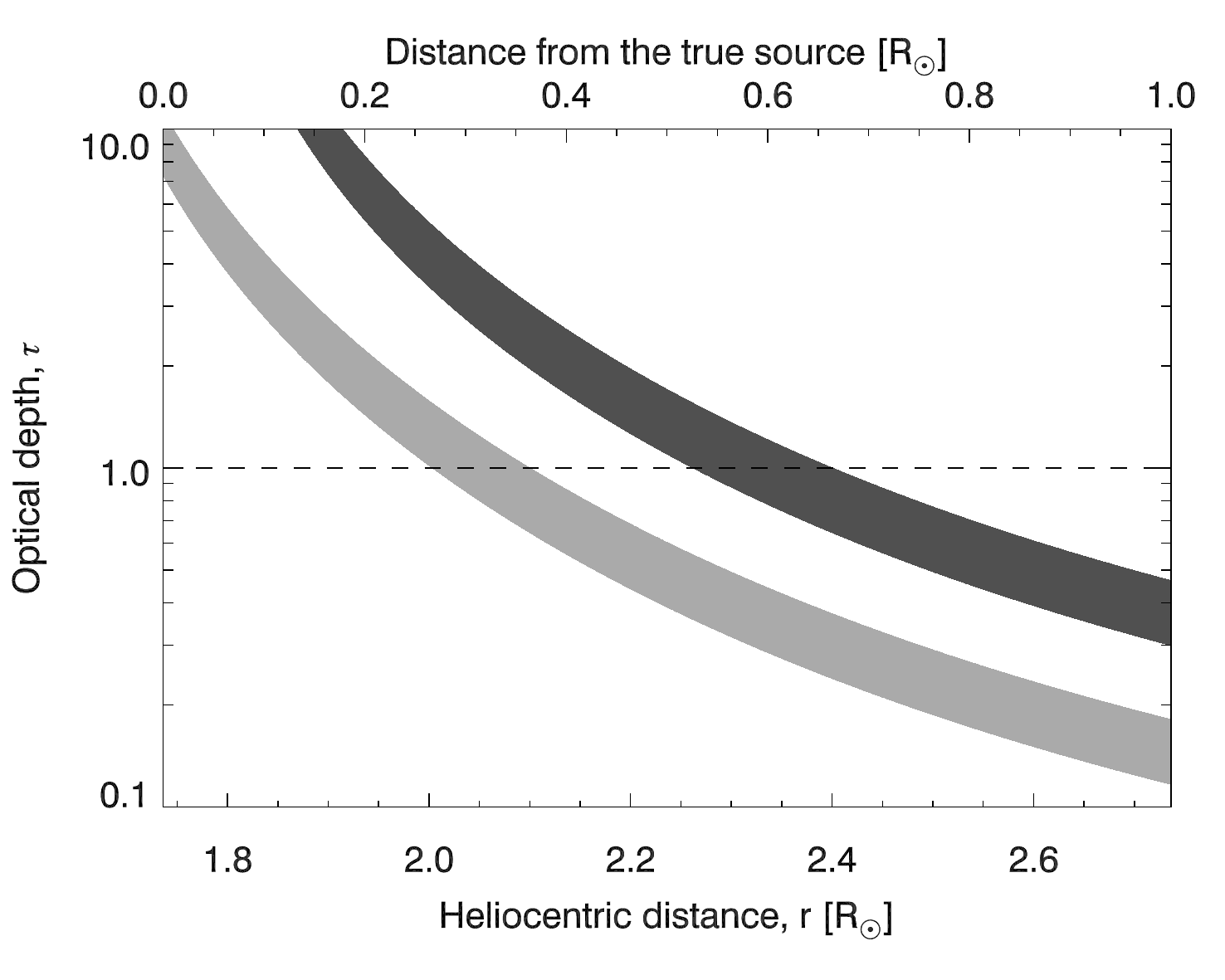}
    \caption{Radio-wave optical depth $\tau$ as a function of heliocentric distance $r$.  The light gray area represents the radio emission at $f_U=40$~MHz for values of $\epsilon^2 / h$ ranging from $4.5\times10^{-5}$ to $7\times10^{-5} \, \mathrm{km^{-1}}$.  The dark gray area represents the radio emission at $f_L=32$~MHz for the same range of $\epsilon^2 / h$.  The dashed line indicates the point at which $\tau =1$, i.e. the transition region from strong to weak scattering from which the radio waves assumed to reach the observer originate.  The heliocentric distance of a source given the 1$\times$Newkirk model is shown (bottom axis), as well as the radial shift that the source will experience from its true location (top axis).
    }
    \label{fig:tau}
\end{figure}

Using previous results (see \citet{1971A&A....10..362S, 1974SoPh...35..153R}) we take $\langle \delta n^2\rangle=\epsilon ^2 n^2$ and the ratio of $\epsilon^2 / h$ as a fixed constant.  Equation (\ref{eqn:dtheta}) shows that the scattering frequency depends on both the density modulation index $\epsilon$ and the characteristic density scale $h$.  Considering the typical values of $\epsilon^2 / h$ ranging from $4.5\times10^{-5}$ to $7\times10^{-5} \, \mathrm{km^{-1}}$ and using the 1$\times$\citet{1961ApJ...133..983N} density model, we calculate the characteristic optical depth for two sources of different frequencies, as illustrated in Figure \ref{fig:tau}.

Assuming that emission appears in the upstream (ahead) $f_L$ and downstream (behind) $f_U$ regions of a shock front, where frequency $f_U > f_L$, the downstream emission will scatter less and therefore the distance where $\tau=1$ is found closer to the Sun.  This is illustrated in Figure \ref{fig:tau} which indicates that 1) the apparent source location is shifted radially outward from the true source location, and 2) the sources produced from the upstream and downstream emission do not appear to be virtually co-spatial as expected, instead, the lower frequency sub-band source is shifted farther away from the higher frequency source.  A source at $f_L = 32$~MHz and a source at $f_U = 40$~MHz were considered for the computation as they represent the average frequency of the lower and upper sub-band sources indicated in Figures \ref{fig:dyn_spec} and \ref{fig:4.5N_vs_obs}.

For the range of $\epsilon^2 / h$ considered, scattering will cause a source observed at 40 MHz to be radially shifted by $\sim 0.3 \, \mathrm{R_\sun}$ from its true location, whereas a source observed at 32 MHz will be shifted by $\sim 0.6 \, \mathrm{R_\sun}$ from its true location (see Figure \ref{fig:tau}).  Consequently, the lower frequency source at $f_L = 32$~MHz is expected to appear as radially shifted by $\Delta R\simeq 0.3 \, \mathrm{R_\sun}$ away from the higher frequency source at $f_U = 40$~MHz.

This radial shifting leads to the fact that the apparent sources, if interpreted as the true sources, will result in an increased apparent density, something that is often observed.  The 1$\times$Newkirk density model (Equation (\ref{eqn:R_model})) can be used to calculate the location $R_i$ of a source at a specific frequency, e.g. at 32 MHz $R_i \simeq 1.74 \, \mathrm{R_\sun}$.  Then the apparent density $n_s$ at the shifted location $R_s = R_i + 0.6 \, \mathrm{R_\sun}$ is $n_e(R_i)/n_e(R_s)=n_e(1.74 \, \mathrm{R_{\sun}})/n_e(2.34 \, \mathrm{R_{\sun}}) \simeq 4.3$ times higher.  Therefore, the source observed at the higher radial distance $R_s$, with respect to the true location $R_i$, causes the density to appear $\sim 4.3$ times larger than what the 1$\times$Newkirk model would predict.  Interestingly, \citet{2018SoPh..293..132M} have reported enhancements of $2.4 - 5.4$ times over the canonical background levels in Type III bursts, which is consistent with our scattering estimates.

\section{Discussion and Conclusions} \label{section:discussion}
We report a LOFAR observation of a split-band Type II radio burst that temporally and spatially relates to a CME event.  The Type II seems to originate from the southern flank of the CME, the origin of which is believed to be near the active region from which an earlier, stronger CME erupted.  This is supported by the fact that coronal dimming is observed below the active region, close to the eruption time of the CME, and that the observed radio sources appear to follow a path away from the location of the dimming.

The source locations of both components of the split-band Type II burst have been determined, for the first time, at the exact same moment in time.  This allows us to study the physical relation of the Type II sub-band sources and make comparisons against the predictions of two widely accepted band-splitting models, specifically the \citet{1974IAUS...57..389S, 1975ApL....16R..23S} and \citet{1983ApJ...267..837H} models.  The average separation between the sources of the upper and lower sub-bands was calculated to be $\sim 0.2 \pm 0.05 \, \mathrm{R_\sun}$.  Such physical separation is larger than what the \citet{1974IAUS...57..389S, 1975ApL....16R..23S} model can explain since the simultaneous emission from the upstream and downstream regions of a shock front should produce nearly co-spatial sources.  We further compare the observed emission source locations to the locations predicted by the standard \citet{1961ApJ...133..983N} density model at equivalent frequencies, and find that the 4.5$\times$Newkirk model is the one that matches the observations best.  As suggested by Figure \ref{fig:4.5N_vs_obs}, both the upper and lower sub-bands are described by the 4.5$\times$Newkirk model.  It should, however, be noted that due to the lower sub-band being observed at the edge of the field-of-view of LOFAR, and perhaps not fully imaged (see Figure \ref{fig:centroid_fit_and_contours}), we are only able to place a lower limit on the apparent heliocentric distance of the lower sub-band sources as they could be underestimated to some degree.  At first glance, the high degree of physical separation between the apparent sub-band sources and the ability to describe the location of both sub-bands using the same density model, leads us to think that the observations support the \citet{1983ApJ...267..837H} explanation of band-splitting.

It is well known that as radio waves transit through the solar corona they suffer from several wave propagation effects, as well as projection effects.  The dominant radio-wave propagation effect in the low corona is believed to be scattering.  Since scattering and projection effects alter the true location of the emission sources, the extent to which they do must be quantitatively evaluated.

A mathematical relation which can be used to estimate the out-of-plane heliocentric distance of the sources of a split-band Type II, as long as they are both imaged at the same moment in time, has been presented.  The assumptions made to obtain this relation are that both sources are found within the same atmosphere (a scenario matching the \citet{1983ApJ...267..837H} model) and that they follow the same path away from the Sun.  Both of these assumptions are supported by the observed source locations (c.f. Figures \ref{fig:centroid_fit_and_contours} and \ref{fig:4.5N_vs_obs}).

\begin{figure}[ht!]
    \centering
    \includegraphics[width=0.75\textwidth, keepaspectratio=true]{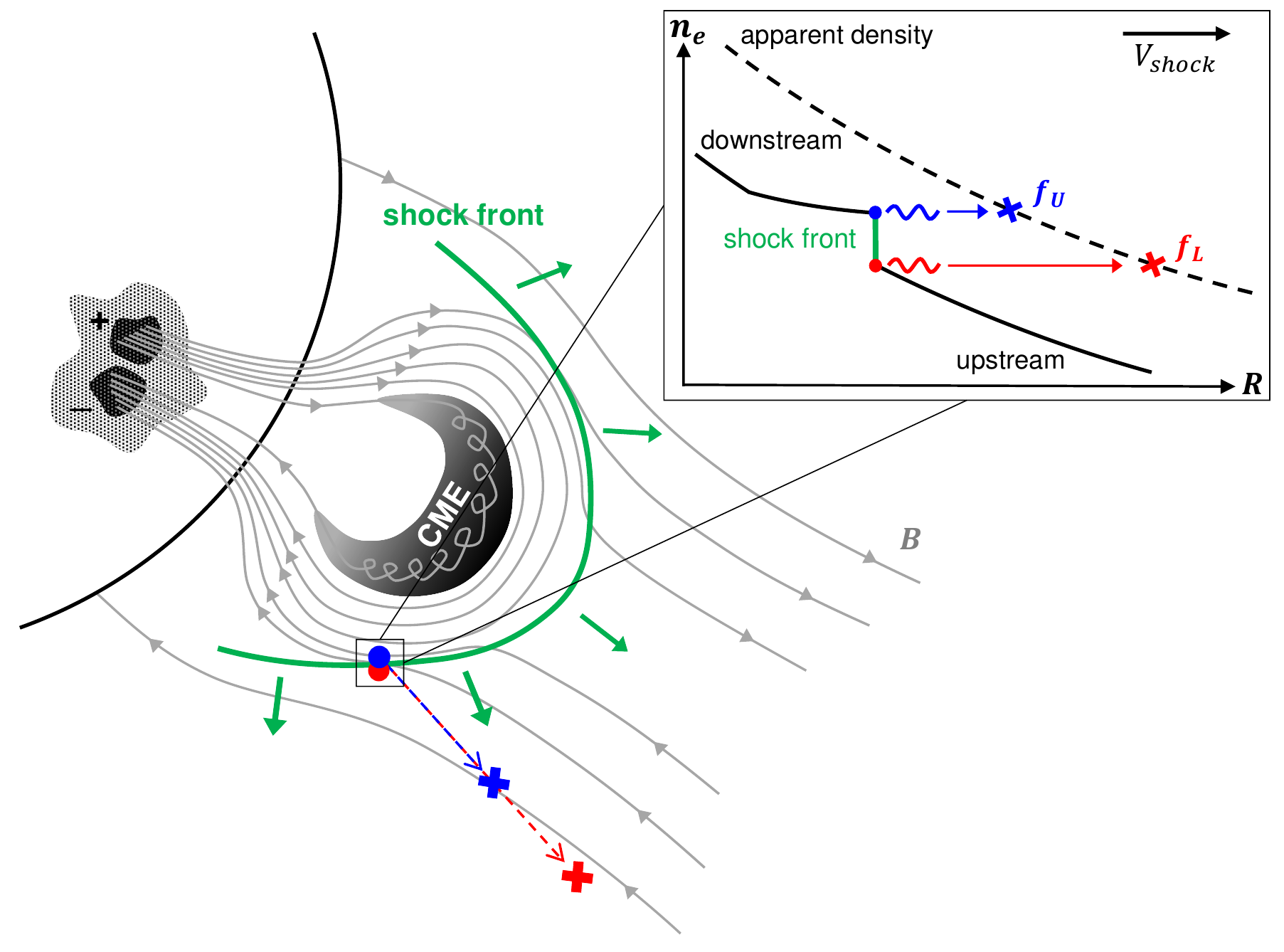}
    \caption{A schematic illustration of the effects that radio-wave scattering has on the true sources of a split-band Type II and the coronal density.  A CME originating from an active region on the solar surface propagates away from the Sun pushing magnetic fields ($B$, solid gray lines) and driving a shock (solid green line).  Two virtually co-spatial emission sources - one for the higher frequency sub-band (blue circle) and one for the lower frequency sub-band (red circle) - form on the two edges of the shock front.  The radiation emitted experiences scattering which causes the sources to shift radially outward from the Sun, as well as appear to be spatially separated (blue and red crosses).  The radial shift also causes the coronal density to appear larger than the true value (black dashed line).
    }
    \label{fig:cartoon}
\end{figure}

An analytical estimation of the effects of scattering has been presented for the first time.  The scattering that split-band Type II radio emissions experience due to small amplitude density fluctuations was shown to result in the following:
\begin{enumerate}
    \item The lower and higher frequency sub-band sources (at 32 and 40 MHz, respectively) are shifted radially farther from the Sun with respect to their true location, making the standard Newkirk density profile of the corona seem larger by a factor of $\sim 4.3$ than its true value.  Thus, the high values of apparent coronal density deduced from the observations (see Figure \ref{fig:4.5N_vs_obs}) can be explained as a side-effect of the scattering experienced by the sources.
    \item The low frequency component of the split band is shifted more than the high frequency component causing a separation of $\sim 0.3 \, \mathrm{R_\sun}$ between the sources at 32 and 40 MHz.  This means that the separation imaged between the sub-band sources ($\sim 0.2 \pm 0.05 \, \mathrm{R_\sun}$) can be explained as an effect of scattering rather than a real physical separation.
\end{enumerate}
The radio-wave scattering effects on the sources of a split-band Type II and the coronal density are presented in a schematic illustration in Figure \ref{fig:cartoon}.  It should be noted that in order to obtain the analytical estimation of the scattering effects, certain assumptions have been adopted which do not reflect the realistic behavior of the solar corona.  More specifically, the ratio of $\epsilon^2 / h$ was kept constant over the range of distances studied.  The degree to which scattering affects the sources will vary according to the number of $\epsilon^2 / h$ assumed.  In this paper, we present the output for a range of values of $\epsilon^2 / h$ (from $4.5\times10^{-5}$ to $7\times10^{-5} \, \mathrm{km^{-1}}$), but it is hard to argue which value is more suitable.  Whilst we are aware of the limitations that such an analytical analysis enforces, it nevertheless emphasizes the importance and extent of scattering effects on radio sources in the corona.

Given the observed distance between the high and low frequency sub-band sources of the studied Type II burst ($\sim 0.2 \pm 0.05 \, \mathrm{R_\sun}$) and by accounting for the scattering effects, we find that the true emission sources could originate from the same spatial location.  This result favors the model proposed by \citet{1974IAUS...57..389S,1975ApL....16R..23S} as an explanation of the band-splitting in Type II bursts since it requires sources that are virtually co-spatial.  To evaluate this conclusion, we allow for the maximum separation between the upper and lower sub-band sources - using the calculated source position errors (see Figure \ref{fig:4.5N_vs_obs}) - and then correct for the estimated radial shifts caused by scattering, i.e. an average of $0.3 \, \mathrm{R_\sun}$ and $0.6 \, \mathrm{R_\sun}$ for the upper and lower sub-bands, respectively.  The remaining physical separation ($\lesssim 0.02 \, \mathrm{R_\sun}$ on average) is insufficient to account for the observed frequency split ($\sim$ 8 MHz) between the two sub-bands, even if very high values of the Newkirk model are assumed.  This is an indication that the \citet{1983ApJ...267..837H} model cannot be used to explain the band-splitting observed after scattering is taken into account.  The density model that best describes the lower frequency sub-band after the corrections for scattering are applied is the 1.3$\times$Newkirk model, whereas the higher frequency sub-band is described by the 1.9$\times$Newkirk model.  Interestingly, the ratio between these two models is a factor of $\sim$ 1.46 which matches the density jump between the upper and lower sub-bands as calculated from the dynamic spectrum ($\sim 1.46$, using the red points of Figure \ref{fig:dyn_spec}), if the \citet{1974IAUS...57..389S,1975ApL....16R..23S} model is invoked (see \citet{2001A&A...377..321V}).

We should, however, note that projection and scattering effects have an opposite impact on the apparent density profile.  The observed heliocentric distance of the sources $R_{obs}$, and the separation between them, depends on the relation $R_{obs} = R_{true} \sin \theta$, where $R_{true}$ is the true source location, and $\theta$ is the projection angle.  When the angle $\theta$ between the observer and the true sources is $90^{\circ}$, the true location of the sources and the true separation between them will be observed.  However, as the angle $\theta$ approaches $0^{\circ}$, the heliocentric distance of the sources will be increasingly underestimated and the separation between the sources of the split band will progressively decrease until no separation can be resolved at $\theta = 0^{\circ}$.  Similarly, the separation between the sub-band sources caused by scattering has been calculated for the radial frame (with respect to the Sun) and can therefore be affected by projection effects.  When two sources of fixed frequencies are imaged in the plane of the sky, the observed separation between them will become progressively smaller as the angle between the radial plane and the plane of the sky increases.

In conclusion, the sources of a split-band Type II burst appear to be spatially separated due to the scattering experienced by the emitted radio waves, although the Type II emission sources can originate from nearly co-spatial regions.  This scenario is consistent with in-situ observations of Type II bursts \citep{1999GeoRL..26.1573B,2000ApJ...544L.163T} where non-thermal electrons are generated in the upstream region but penetrate into the downstream region creating Langmuir waves and split-band radio emission.  Extrapolations of the higher and lower frequency sub-bands of Type II bursts were found to match the density jumps recorded at 1 AU \citep{2001A&A...377..321V, 2002ESASP.506..335M}, indirectly favoring the \citet{1974IAUS...57..389S, 1975ApL....16R..23S} model.

The importance of radio-wave scattering effects on the source locations of split-band Type II bursts is highlighted.  However, one should be mindful of other effects that - even if not dominant - will affect the radio sources.  Besides projection effects, refraction of the radio waves will also have opposite consequences to scattering on the apparent sources as it can shift them radially closer to the Sun.  The complete picture of interactions disrupting the radio waves needs to be understood before any reliable inferences can be made regarding the local coronal conditions from either spectroscopic or imaging observations of radio sources.  Considering the effects of scattering, our results distinguish between band-splitting interpretations that require the true emission sources to be virtually co-spatial and interpretations that require the true sources to be spatially separated.  As a consequence, we have provided supporting evidence for band-splitting interpretations such as that proposed by \citet{1974IAUS...57..389S, 1975ApL....16R..23S} which explains the observed band-splitting as radiation emitted from the upstream and downstream parts of a shock front, two nearly co-spatial regions.

\acknowledgments
N.C. and E.P.K. were supported by an STFC studentship and the STFC consolidated grant ST/P000533/1.
M.T. acknowledges the support by the FFG/ASAP Program under grant no. 859729 (SWAMI).
The authors acknowledge the support by the international team grant (\url{http://www.issibern.ch/teams/lofar/}) from ISSI Bern, Switzerland.
This paper is based (in part) on data obtained from facilities of the International LOFAR Telescope (ILT) under project code LC3\_012 and LC4\_016. LOFAR \citep{2013A&A...556A...2V} is the Low Frequency Array designed and constructed by ASTRON.  It has observing, data processing, and data storage facilities in several countries, that are owned by various parties (each with their own funding sources), and that are collectively operated by the ILT foundation under a joint scientific policy.  The ILT resources have benefitted from the following recent major funding sources: CNRS-INSU, Observatoire de Paris and Universit\'e d'Orl\'eans, France; BMBF, MIWF-NRW, MPG, Germany; Science Foundation Ireland (SFI), Department of Business, Enterprise and Innovation (DBEI), Ireland; NWO, The Netherlands; The Science and Technology Facilities Council, UK; Ministry of Science and Higher Education, Poland.
The SOHO/LASCO data used here are produced by a consortium of the Naval Research Laboratory (USA), Max-Planck-Institut fuer Aeronomie (Germany), Laboratoire d'Astronomie (France), and the University of Birmingham (UK). SOHO is a project of international cooperation between ESA and NASA.
The authors would also like to thank the SDO team for the data.

\bibliography{references}

\end{document}